\documentclass[amssymb,aps,prl,twocolumn,superscriptaddress,showpacs]{revtex4}

\usepackage{graphicx}

\usepackage[latin1]{inputenc}

\begin{document}

\title{Filament depolymerization by motor molecules}

\author {Gernot  A. Klein}
\affiliation{Max Planck Institute for Physics of Complex Systems, D-01187 Dresden, Germany}

\author{Karsten Kruse}
\affiliation{Max Planck Institute for Physics of Complex Systems, D-01187 Dresden, Germany}

\author{Gianaurelio Cuniberti}
\affiliation{Max Planck Institute for Physics of Complex Systems, D-01187 Dresden, Germany}
\affiliation{Institute for Theoretical Physics, University of Regensburg,
D-93040 Regensburg, Germany}

\author{Frank J\"ulicher}
\affiliation{Max Planck Institute for Physics of Complex Systems, D-01187 Dresden, Germany}

\date{\today }


\begin{abstract}
Motor proteins that specifically interact with the ends of cytoskeletal filaments can induce 
filament depolymerization. A phenomenological description of this 
process is presented. We show that under certain conditions motors
dynamically accumulate at the filament ends. We compare simulations of two microscopic
models to the phenomenological description. The depolymerization rate can exhibit maxima and dynamic 
instabilities as a function of the bulk motor density for processive depolymerization.
We discuss our results in relation to experimental studies of Kin-13 familiy motor proteins.
\end{abstract}

\maketitle

Many active processes in cells are driven by highly specialized 
motor proteins which interact with filaments of the cytoskeleton.
Important examples are cell locomotion, cell division and the transport
of organelles inside the cell \cite{albe02}.
Cytoskeletal filaments are linear aggregates of proteins, for example actin and tubulin. 
Actin filaments and microtubules are dynamic and can rapidly change their lengths
by addition and removal
of subunits at the ends \cite{albe02,pesk93,dogt93}. 
Filaments show a structural asymmetry, which provides
a direction for motion and force generation of bound molecular motors.
These proteins are able to transduce the chemical energy
of a fuel which is ATP, to mechanical work while interacting with
a filament \cite{howa02,juli97}.

In addition to generating forces along filaments, motors can also
interact with filament ends where they may
influence the polymerization rate and thus the filament length. Examples are provided by the 
members of the Kin-13 subfamily of kinesin motor proteins \cite{desa99,hunt03,brin04}. A particular
example is the mitotic centromere-associated kinesin (MCAK) which regulates the length 
of microtubules during cell division \cite{word95}.
In the course of cell division, the chromosome pairs are separated by the mitotic spindle. 
In this process, shortening microtubules generate forces pulling the
chromosomes towards the opposing poles of the cell. MCAK is localized
at the microtubule ends which interact with chromosomes \cite{word95} 
and it has been shown that it induces depolymerization of microtubules \cite {hunt03}. 
In vitro assays and single molecule studies have shown that MCAK accumulates
at both ends of stabilized microtubules and induces depolymerization at a
rate which depends on the bulk motor concentration while at the same time
MCAK molecules do not generate directed motion along microtubules \cite{hunt03}.

In this paper, we discuss the dynamics of 
motor molecules which induce the shortening of the ends of filaments to which they
bind using both a phenomenological description and more microscopic models.
For simplicity, we consider one filament end and use a semi-infinite geometry. 
The density of bound motors at a distance $x\ge0$ from the depolymerizing filament
end is denoted $\rho(x)$. Here, we use a reference frame 
in which the depolymerizing end is located
at $x=0$ for all times.  Motors occur in the bulk solution at concentration $c$. They 
bind to and detach from filaments
with rates $\omega_{\rm a} c/\rho_{\rm max}$ and $\omega_{\rm d}$, respectively,
where $\rho_{\rm max}$ is the maximal density of motors 
for which binding sites on the filament saturate. 
Bound motors diffuse along the filament with a diffusion coefficient $D$
and may also exhibit a directed average motion with velocity $v_0$.
Note, that $v_0$ in general depends on the density 
of motors $\rho$ \cite{lipo01,krus02,parm03,klum04}.
The density profile along the
filament then obeys
\begin{eqnarray}
\label{mfa}
\partial_{t} \rho + \partial_{x}j &=& \omega_{\rm a} c \left(1-\frac{\rho}{\rho_{\rm max}} \right) - 
\omega_{\rm d} \rho \quad .
\end{eqnarray}
The current of motors is given by $j=-D \partial_{x} \rho - v \rho$.
Here, $v=v_0+v_{\rm d}$ is the total velocity of motors with respect to the filament end, 
with $v_{\rm d}\ge0$ denoting the depolymerization velocity.
It is related to the rate $\Omega$ of subunit removal from the end 
by $v_{\rm d}=\Omega a/N$, where $a$ is the size of a subunit and $N$ 
the number of protofilaments in the filament.

We assume that the rate of filament depolymerization is regulated by motors
bound to the filament end. Therefore,
the rate of subunit removal is a function $\Omega(\rho_0)$ of the 
motor density $\rho_0=\rho(x=0)$ at the end. It is useful 
to systematically expand $\Omega$
in powers of $\rho_0$
\begin{eqnarray}
\label{vd}
\Omega(\rho_{0}) &=& \Omega_0 + \Omega_1 \rho_{0} + \Omega_2 \rho_{0}^{2} + \mathcal{O}(\rho_0^3) \quad.
\end{eqnarray}
Here, we have introduced the expansion coefficients $\Omega_i$.
The subunit removal rate in the absence of motors 
$\Omega_0$ in general depends on buffer conditions. In situations where filaments are stabilized,
$\Omega_0=0$.  For motors which induce filament depolymerization,
$\Omega_1>0$. Since  the rate $\Omega$ saturates for large densities, typically 
$\Omega_2 < 0$.

The description is completed by specifying the boundary conditions at 
$x=0$ and for $x\to\infty$. At $x=0$ the current $j(x=0)$ 
at the filament end
equals the net rate $J$ at which motors attach to the filament end.
Since motors attached to the end induce depolymerization, the rate
$J(\rho_{0})$ is a function of the motor density $\rho_0$ at the end 
and also depends on buffer conditions. 
Again, we express $J$ by an expansion in powers of $\rho_{0}$:
\begin{equation}
\label{jnull}
J(\rho_{0})=J_{0}+J_{1} \rho_{0}+J_{2} \rho_{0}^{2} +\mathcal{O}(\rho_0^3) \quad,
\end{equation}
Here, $J_{0}$ is the rate of direct motor attachments to the end. The coefficients $J_{1}$ and $J_{2}$ characterize how interactions between motors and the filament end influence the
detachment rate of motors.
If $v > 0$, motors typically detach from the end 
and thus $J<0$. 

Finally, for large $x$ we require 
that the density $\rho$ approaches the equilibrium value of the 
attachment-detachment dynamics
\begin{eqnarray}
\label{rhoinf}
\rho_{\infty} &=& \frac{\omega_{\rm a} c \rho_{\rm max}}{\omega_{\rm a} c+\omega_{\rm d} \rho_{\rm max}}
\end{eqnarray}

If a motor bound to the end removes a filament subunit, it may fall off the filament with
this subunit  or it may stay bound to the filament. 
The tendency of a motor to stay attached while removing subunits
can be described by its processivity. 
In our phenomenological description, we define the effective processivity
\begin{equation}
p_{\rm eff} = 1-\left\vert \frac{J(\rho_{0})-J(0)}{\Omega(\rho_{0})-\Omega(0)}\right\vert \quad .
\end{equation}
It differs from the processivity of an isolated motor due to
collective effects resulting from interactions between motors at the end and thus
depends on $\rho_0$.
If $p_{\rm eff} \le 0$, more motors detach from
the end than subunits and the motor-induced depolymerization is
non-processive. However, if $0 < p_{\rm eff} \le 1$, a given motor can remove
more than one subunit\footnote{Note that the motor density at the end cannot 
exceed the maximal density,  $\rho_{0}\le\rho_{\rm max}=N/a$. Thus, the effective 
processivity vanishes for $\rho_{0}=\rho_c $ with $\rho_c \leq \rho_{\rm max}$.}.

\begin{figure}[t]
\begin{center}
\includegraphics[scale=0.25]{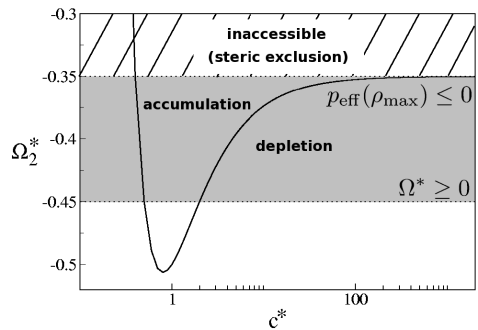}
\caption{Regimes of accumulation and depletion of motors 
at the shrinking filament end as a function of 
the coefficient $\Omega_{2}^{*}=\Omega_{2} \rho_{\rm max}/(D \omega_{\rm d})^{1/2}$ and 
the bulk motor concentration $c^{*} = \omega_{\rm a} c /\omega_{\rm d} \rho_{\rm max}$ for $J_{0}= \Omega_{0}=J_{2}=0$ and 
$J_{1}/(D \omega_{\rm d})^{1/2}=-0.1, \Omega_{1}/(D \omega_{\rm d})^{1/2}=0.5$. Accumulation occurs above the solid line, depletion occurs below. 
The grey area indicates the region of physical interest. Above this area, $p_{\rm eff}(\rho_{\rm max})>0$
which is forbidden by steric exclusion of particles. Below this area, filaments polymerize at high
motor concentration.
}
\label{figphase}
\end{center}
\end{figure}

For simplicity, we ignore the density dependence of $v_0$.
In this case, Eq.~(\ref{mfa}) approaches
for large times $t$,  a steady state given by $\rho = \rho_{\infty} + \left ( \rho_{0} - \rho_{\infty} \right ) \exp [- x/ \lambda]$
The characteristic length is
\begin{eqnarray}
\label{decay}
\lambda = \frac{2 D}{v+\left [v^{2} + 4D ( \omega_{\rm a} c/\rho_{\rm max} +\omega_{\rm d} )\right ]^{1/2}}  
\;.
\end{eqnarray}
The steady state value of $\rho_0$ is obtained by inserting this expression
in Eq. (\ref{jnull}). 
In the following, we consider the case where $v_{\rm d} \gg |v_{0}|$ and
the spontaneous velocity $v_0$ can be neglected. Indeed, 
experimental observations of MCAK show that $v_{\rm d} \gg |v_{0}|$
\cite{hunt03}.
Depending on parameters, motors either accumulate or deplete
at the filament end, see Fig.~\ref{figphase}.
Accumulation at the end occurs for $\Omega_{2}>\Omega_{2}^{\rm (a)}$, where
\begin{eqnarray}
\label{accumulation}
\Omega_{2}^{\rm (a)} &=& - \frac{\Omega_{1}+ J_{2} \rho_{\rm max}}{ \rho_{\infty}} - \frac{ \Omega_{0}+J_{1} \rho_{\rm max}}{\rho_{\infty}^{2}} - \frac{J_{0} \rho_{\rm max}}{\rho_{\infty}^{3}}.
\end{eqnarray}
Motor accumulation can exhibit a reentrant behavior
as a function of increasing bulk motor concentration, see Fig.~\ref{figphase}.

Our phenomenological description reveals that motors can dynamically accumulate at the
filament end even if their binding affinity to the
end is not larger than in the bulk, $\Omega_a=0$. 
In this process,  motors which bind along the filament are  subsequently captured by the 
retracting filament end. Dynamical accumulation of motors is a collective phenomenon and 
requires a sufficiently large effective processivity. 

\begin{figure}[t]
\begin{center}
\includegraphics[scale=0.30]{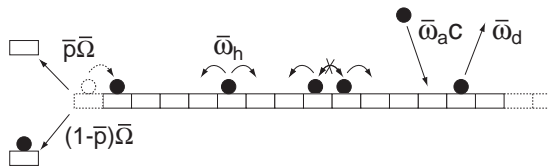}
\caption{Discrete model of motor-induced
filament depolymerization. 
Motors attach to empty sites at 
rate $\bar{\omega}_{\rm a} c$ and detach with rate 
$\bar{\omega}_{\rm d}$. The hopping rate to free neighbouring sites 
is $\bar{\omega}_{\rm h}$. Occupied sites are removed from the end with 
rate $\bar{\Omega}$. With probability $\bar{p}$, the particle remains attached
to the end when a subunit is removed.}
\label{fighopping}
\end{center}
\end{figure}

In order to obtain a physical picture of the microscopic events which influence the
effective processivity as a result of crowding, we extend discrete
stochastic models for motor displacements along filaments
\cite{lipo01,krus02,parm03,klum04} to capture subunit removal at the ends.
Motors are represented by
particles which occupy discrete binding sites indexed by $i=1,2,3,\dots$
arranged linearly on a filament which consists of a single proto-filament (N=1), 
see Fig.~\ref{fighopping}.  Here, $i=1$ denotes the binding 
site at the filament end. Each site is either empty ($n_i=0$) or 
occupied ($n_i=1$). Particles move stochastically to neighboring empty 
sites with rate $\bar{\omega}_{\rm h}$ in both directions. Here, 
we assume  again that $v_0$ can be neglected as compared to $v_{\rm d}$. 
In addition, particles attach to and detach from the lattice with rates 
$\bar{\omega}_{\rm a}c$ and $\bar{\omega}_{\rm d}$, respectively. 
The rates of particle attachment and detachment at the end site 
$i=1$ differ from the bulk rates and are denoted 
$\bar{\Omega}_{\rm a} c$ and $ \bar{\Omega}_{\rm d}$. 

We first consider the situation where only one 
particle is present. If this particle is not bound at the 
end, the end is stable. If the particle is bound at $i=1$, this subunit 
is removed from the filament with rate $\bar{\Omega}$. This
process can occur in two different ways: (i) with probability $\bar p$
the particle stays bound to the new filament end after the first 
subunit is removed; (ii) with probability $1-\bar p$ the particle detaches 
from the filament together with the removed subunit.
If $\bar p$ is close to one, a single particle is processive and can repeatedly remove
subunits from the end without falling off. 

Processive removal of a subunit requires simultaneous
interaction of a motor with the end and the adjacent subunit. Therefore, this process is 
affected by the presence of other particles bound to the filament near 
the end. In particular, if site $i=2$ is occupied while subunit $i=1$ is 
removed, the new end site is already occupied and the 
particle at $i=1$ cannot stay attached. 
We distinguish two cases with different behaviors in this situation.
Model A describes the case where subunit removal by a motor requires an empty adjacent
binding  site,
while model B corresponds to the case where the rate of subunit removal
is independent of the occupation of the adjacent site.
In model A, the probability per unit time to remove the end if $n_1=1$ is
$\bar\Omega^A=\bar \Omega (1-\bar p n_2)$. Here we assume that
the processivity characterized by $\bar p$ is unaffected by the 
occupation of the neighboring site. Crowding at the end 
obstructs cutting and reduces the rate of subunit removal. 
In model B, an occupied adjacent site will reduce the processivity 
of a motor but does not affect the depolymerization rate, 
i.e., $\bar\Omega^B=\bar\Omega$. 

We can represent
the dynamics of the system by a Master equation
for the probability $P\{n_i,t\}$ to find a configuration of 
lattice occupation $(n_1,n_2,..)$ at time $t$. This leads to 
expressions for the rate of change of average 
occupation numbers valid for $i\geq 2$:
\begin{eqnarray}
\label{hoppingbulk}
\frac{d \langle n_{i} \rangle}{dt} &=& \bar{\omega}_{\rm h} \left ( \langle n_{i+1} \rangle - 2 \langle n_{i} \rangle +  \langle n_{i-1} \rangle \right )
  + \bar{\omega}_{\rm a} c\langle 1-n_{i} \rangle\nonumber \\
  & -& \bar{\omega}_{\rm d} \langle n_{i} \rangle 
  +  \langle \bar{\Omega}^{A,B}n_{1}  \left (n_{i+1}-n_{i} \right ) \rangle\;.
\end{eqnarray}
At the filament end, $i=1$,
\begin{eqnarray}
\label{hoppingedge}
\frac{d \langle n_{1} \rangle}{dt} &=& \bar{\omega}_{\rm h} \langle n_{2} - n_{1} \rangle 
+ \bar{\Omega}_{\rm a} c\langle 1-n_{1} \rangle - \bar{\Omega}_{\rm d} \langle n_{1} \rangle \nonumber\\
&   - & (1-\bar{p})\bar{\Omega}\langle  n_{1}(1-n_2) \rangle \;. 
\end{eqnarray}
Using a mean-field approximation, replacing two-point correlators
$\langle n_{i}n_{i+1} \rangle$ by $ \langle n_{i} \rangle \langle 
n_{i+1} \rangle$, we obtain from Eqs.~(\ref{hoppingbulk}) and 
(\ref{hoppingedge}) differential equations and boundary conditions 
identical to Eqs. (\ref{mfa})-(\ref{jnull}) with $\rho(x=a (i-1))= \langle n_i 
\rangle/a$. 
This procedure leads to explicit expressions for the values of the coefficients $\Omega_i$
and $J_i$ introduced above. 
For both model A and B we find, using this approximation, $D=a^2\omega_h$, $\Omega_{1} = a \bar{\Omega}$, $J_{0} 
= \bar{\Omega}_{\rm a} c$ and $J_{1} = -a (\bar{\Omega}_{\rm a} c+ \bar{\Omega}_{\rm d} +  (1-\bar{p}) \bar{\Omega})$.
The nonlinear coefficients $\Omega_{2}$ and $J_{2}$ are model dependent. In model
A, $\Omega_{2} = - a^{2} \bar{p} \bar{\Omega}$ and $J_{2} = 0$ whereas
for model B, $\Omega_{2}= 0$ and $J_{2}=-a^{2} \bar{p} \bar{\Omega}$. In both models all higher
order coefficients $\Omega_n$ and $J_n$ vanish. 

\begin{figure}[t]
\begin{center}
\includegraphics[scale=0.3]{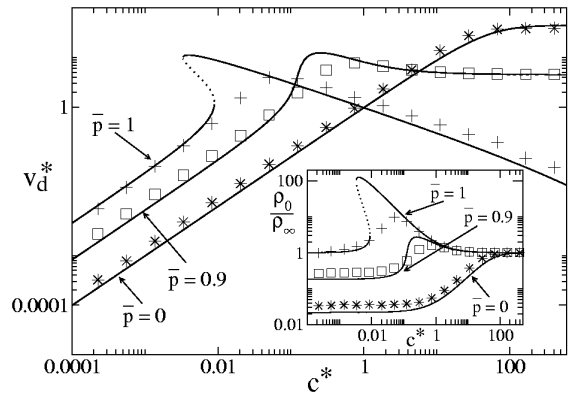}
\caption{The velocity $v_{\rm d}^*=v_{\rm d} /a(\bar{\omega}_{\rm h} \bar{\omega}_{d})^{1/2}$ of depolymerization as a function of the
bulk motor concentration $c^*=\bar{\omega}_{\rm a} c/\bar{\omega}_{\rm d}$ obtained in 
simulations of model A for different values of the processivity $\bar p=0$, $0.9$ and $1$ (symbols). 
For comparison the corresponding solutions of the phenomenological equations are displayed (lines). Inset: The accumulation of motors, characterized by the ratio $\rho_0/\rho_{\infty}$ of motor density 
at the end and far from the end is shown as a function of $c^*$ for the same situations. Parameter 
values are $\bar{\omega}_{\rm a}=\bar{\Omega}_{\rm a},  \bar{\omega}_{\rm d}=\bar{\Omega}_{\rm d}=
0.008\,\bar{\omega}_{\rm h}$ and $\bar{\Omega}=4\,\bar{\omega}_{\rm h}$.}
\label{mfasim}
\end{center}
\end{figure}

Fig.~\ref{mfasim} displays the depolymerization velocity $v_{\rm d}$  obtained in the mean field
theory corresponding to model A as a function of the bulk monomer concentration $c$  
for different values of $\bar p$. For large $c$ the velocity
saturates at $v_{\rm d}(\rho_{\rm max})$, while it increases linearly for small $c$. For sufficiently large 
$\bar p$, the velocity $v_{\rm d}$ exhibits a maximum as a function of $c$. Increasing $\bar p$ further, a dynamic 
instability appears where two stable states with different $v_{\rm d}$ coexist within a range of $c$ values. A third unstable state is indicated by a broken line.
Results of stochastic simulations of model A are shown for comparison.
Mean field theory and stochastic simulation agree quantitatively except in the vicinity of the 
dynamic instability. 
Fluctuations conceal the dynamic instability present in mean field theory. 
The inset to Fig.~\ref{mfasim} shows the relative accumulation $\rho_0/\rho_{\infty}$ of motors.
For sufficiently large $\bar p$, motors accumulate as $c$ is increased. We note that in model B,
no dynamic accumulation of motors occurs and $\rho_0<\rho_\infty$.

In summary, 
we have shown that a positive
effective processivity $p_{\rm eff}$ of subunit removal is essential to achieve 
dynamic accumulation of motors at the filament end. This effective processivity is a collective effect %
and results from steric exclusion of motors bound near the end.  
The phenomenological description given by Eqs.~(\ref{mfa})-(\ref{jnull}) is general and valid
irrespective of details of the mechanism of motor induced subunit removal 
and of the structure
of the depolymerizing filament end.
We have restricted ourselves to effects corresponding to the lowest order terms in the expansions
of Eqs. (\ref{vd}) and (\ref{jnull}). While higher order terms could lead to additional effects, our simulations
of microscopic models indicate that 
these terms are unimportant (see Fig. 3). We have focussed on
excluded volume effects at the filament end and  have considered the case $v_0=0$ where some of these 
effects in the bulk disappear. For $v_0\neq 0$,
the interplay between bulk and end excluded volume effects could lead to new phenomena which will be subject
of future work. 

The stochastic 
models A and B provide a physical picture of the 
cooperativity and processivity
of motors bound at the end of a single protofilament.  
Interactions between motors lead to different rates of subunit removal 
in the two models. In our stochastic simulations for $N=1$, the dynamic instability
of steady states which is found in the mean field analysis is concealed 
by fluctuations. We expect that for larger numbers of protofilaments this effect of
fluctuations is reduced. Therefore a signature of a dynamic instability could reappear
for microtubule depolymerization leading to bistability and switch like
changes of depolymerization velocities. In the mitotic spindle this instability could be relevant
for chromosome oscillations, which have been observed \cite{skib93}.

Accumulation of motors at the filament end described by Eq. (9) can
occur  as a result of three different mechanisms.  Motors can
accumulate by directly binding to the filament end  
if they have a higher affinity to the end than to subunits along the
filament. This effect dominates if the
total velocity $v=v_0+v_{\rm d}$ is small, 
$v^2\ll 4D(\omega_{\rm a} c/\rho_{\rm max}+\omega_{\rm d})$. In this case, 
the localization length is given by the diffusion length during the attachment time 
$\lambda\simeq D^{1/2}(\omega_{\rm a} c/\rho_{\rm max} +\omega_{\rm d})^{-1/2}$.
A second mechanism of accumulation is given by transport of motors 
to the end with velocity $v_0\gg v_{\rm d}$.
In the third case, motors that bind along the filament 
are captured by the shortening end. 
This dynamic accumulation 
mechanism dominates for $v_0\ll v_{\rm d}$ and 
$v_{\rm d}^2\gg 4D(\omega_{\rm a} c/\rho_{\rm max}+\omega_{\rm d})$. The 
localization length is $\lambda\simeq D/v_{\rm d}$. 
The first and last cases can lead to accumulation at both ends of a filament
of finite length, while in the second
case motors accumulate at one end only. 

Our results can be related to experiments on members of the Kin-13
family of kinesins.  The depolymerization velocity $v_d$ as a function of bulk
motor concentration has been measured for MCAK and it has been shown that
MCAK accumulates at both ends \cite{hunt03}. 
The observed velocity $v_d$  is consistent with both models A and B since it  does not exclude
the possibility of a maximal velocity for intermediate motor concentrations.
Accumulation at the end suggests that a mechanism similar to
model A is more likely to be at work. Indeed, experiments indicate that a collection of MCAK motors
processively depolymerize microtubules \cite{hunt03} consistent with model A.
Present data cannot rule out a mechanism akin to model B where motor
accumulation is still possible if the affinity of motors to the filament end is high.
In future experiments, model B could be ruled out if a 
a maximum of the depolymerization 
velocity at intermediate motor concentration would be observed as suggested by our theory. The members XKCM1 and XKIF2 of the Kin-13 family,
can depolymerize microtubules with or without
accumulation of motors at the end, depending on the conditions under which microtubules
have been stabilized \cite{desa99}. 
Furthermore, it has been suggested that processivity is reduced under conditions where motors do not
accumulate \cite{desa99}. 
Thus, stabilization of microtubules could influence the microscopic mechanisms
of collective subunit removal by a change
in the microtubule lattice structure, leading to reduced processivity $\bar p$ or
a mechanism similar to model B.
 
The theory developed here is not restricted to  motors of the Kin-13 family which interact with microtubules
but applies in general to  associated proteins which regulate the dynamics of filament ends.
Actin depolymerization by ADF/cofilin as well as the polymerization of actin by formin are further
examples of such processes \cite{pruy02,sago02}. In addition to the conventional action of motor proteins, 
filament polymerization and depolymerization by processively acting end-binding
proteins are expected to play a key role in cytoskeletal dynamics and self-organization.
 
We thank Stefan Diez and Joe Howard for fruitful discussions and stimulating collaboration. G.C. acknowledges 
support by the Volkswagen Stiftung.

\end{document}